\documentclass[aps,prl,twocolumn,showpacs,floatfix,showkeys,groupedaddress]
{revtex4}
\usepackage{graphicx}
\usepackage{dcolumn}
\usepackage{bm}
\begin{document}
\preprint{PRL}
\title{Supersymmetric Dark Matter and the Extragalactic Gamma Ray
Background}
\author{Dominik Els$\mathrm{\ddot{a}}$sser}
\email{elsaesser@astro.uni-wuerzburg.de}
\author{Karl Mannheim}
\affiliation{Institut f$\ddot{u}$r Theoretische Physik und Astrophysik,
Universit$\ddot{a}$t W$\ddot{u}$rzburg}
\date{30.03.2005}
\begin{abstract}
We trace the origin of the newly determined extragalactic gamma ray
background from EGRET data to an unresolved population of blazars and
neutralino annihilation in cold dark matter halos. Using results of
high-resolution simulations of cosmic structure formation, we calculate
composite spectra and compare with the EGRET data. The resulting best-fit
value for the neutralino mass is $\mathrm{m_{\chi}=515_{-75}^{+110}~GeV}$ 
(systematic errors $\mathrm{\sim30\%}$).
\end{abstract}
\maketitle
The origin of the extragalactic gamma-ray background (EGB) has been
discussed since the seminal paper on gamma ray astrophysics by Morrison in
1958 \cite{Morrison}. Diffuse, isotropic gamma-ray background radiation
results either from the emission of numerous sources too faint to be
resolved, or from weakly interacting massive particles ("WIMPs") that have
survived as a fossil record of the early Universe. The EGRET spark chamber
detector on board the Compton Gamma Ray Observatory completed an all-sky
survey above 30~MeV, collecting data from 1991 until 2000 \cite{CGRO}. 
Subtraction of the foreground plays an equally important role in
determining the extragalactic gamma-ray background as in the case of other
cosmological precision measurements, e.g. the measurements of the microwave
background and its anisotropies. The discovery of a residual galactic gamma
ray halo at GeV-energies \cite{Dixon} prompted improvements of the
foreground model used in the analysis of the EGRET data. A new
determination of the intensity of the EGB in the energy range of 30~MeV --
50~GeV has been accomplished using the numerical code GALPROP for modeling
the galactic gamma-ray foreground, now including an Inverse-Compton (IC)
component \cite{Strong}. The EGB spectrum has two components: a
steep-spectrum power law with index $\mathrm{\alpha = -2.33}$ and a strong
bump at a few GeV. The first analysis of the EGB \cite{Sreekumar} did not
reveal as clearly this spectral structure. Guided by the observation that
the net spectral index of $-2.10\pm 0.03$ was tantalizingly close to the
mean spectral index of the resolved extragalactic EGRET sources (all but
Centaurus A and the Large Magellanic Cloud are blazars), it was then
concluded that faint, unresolved blazars were responsible for up to 25\% or
100\% of the background, respectively 
\cite{ChiangMukher,SteckerandSalamon}. Physically related sources (such as
radio galaxies), large-scale structures \cite{Blasi} or gamma ray bursts
\cite{Totani} could also contribute to the EGB. Whatever astrophysical
scenario may be considered, however, a universal multi-GeV bump resulting
from the superposition of the spectra of a large, diverse population of
sources remains suspicious.\\ 
By contrast, the observed energies of the excess bump appear naturally in
the context of models involving weakly interacting, annihilating cold
dark matter. MeV-scale dark matter particles have recently been discussed
as a source of the galactic positronium halo \cite{Boehm}. The Lee-Weinberg
criterion for thermal freeze-out during the hot Big Bang, however, renders
weakly interacting particles with masses much larger than that of the
proton natural candidates for the cold dark matter \cite{Primack}.
Independently, supersymmetry calls for a new stable particle with weak
interactions and a mass scale close to $\mathrm{E_{\rm
F}=(1/\sqrt{2}G_{F})^{1/2}\simeq 246}$~GeV, probably the lightest
neutralino ($\mathrm{\chi_{1}^{0}}$) \cite{SUSY}. The annihilation of 
the Majorana neutralinos in dark
matter halos - starting from the freeze-out in the hot Big Bang and
continuing until the present day - produces electromagnetic radiation
(along with $\mathrm{\nu}$, $\mathrm{p}$, $\mathrm{e}$) from the decay
chains of short-lived heavy leptons or quarks. Annihilation lines
\cite{BEU} would only arise from the loop-level processes
$\mathrm{\chi\chi\rightarrow \gamma\gamma\;and\;\chi\chi\rightarrow
Z^{0}\gamma}$, and thus their intensities are generally expected to be
rather small. The continuum gamma-ray energies are kinematically lowered by
factors of the order of ten \cite{Gondolo}. Obviously, the number of dark
matter halos must be much larger than any possible astrophysical gamma-ray
source population, and hence the main signature of cosmological neutralino
annihilation should actually be a rather narrow bump in the EGB at about
10~GeV \cite{Elsaesser1}.
In this Letter, we show that the observed bump in the EGB could well be
this signature of dark matter annihilation, and that there exists an
allowed range of neutralino candidates in the cosmologically constrained
MSSM naturally explaining this feature when combined with a steep
astrophysical power-law spectrum component. 
\section{Modeling the Annihilation Component} With the differential gamma
ray energy distribution from jet fragmentation and $\mathrm{\pi^{0}}$-decay
df, observed energy E and redshift z, the extragalactic gamma ray intensity
due to WIMP annihilation can be written as
$\mathrm{\Phi_{\gamma}\,(E)=c/4\pi H_{0}\times 1/2\left\langle \sigma
v\right\rangle_{\chi}
\;\Omega^{2}_{DM}\;\rho^{2}_{crit}/m_{\chi}^{2}\times}$\\$\mathrm{\int_{0}^
{z_{max}}[\left(1+z\right)^{3}\,\kappa (E,z)\,\Gamma\left(z\right)\,\Psi \,
df_{E(1+z)}]/}$$\mathrm{\xi\left(z\right)\,dz\;}$,\\
\cite{BEU,cl,Elsaesser1} where $\mathrm{\rho_{crit}}$ is the critical
density. Since we consider contributions from annihilations at high
redshifts, gamma-ray absorption is included via the attenuation function
$\mathrm{\kappa (E,z)}$. For $\mathrm{0\leq z\leq 5}$ we use the
attenuation derived from star formation history \cite{Kneiske}, whereas for
$\mathrm{z>5}$ the absorption from interactions with the
cosmological relic radiation field \cite{Zdziarski} is employed. The range
of integration is limited to $\mathrm{0\leq z\leq 20}$; gamma rays from
higher redshifts are negligible. The parameter $\mathrm{\xi\left(z\right)}$
is given by
$\mathrm{\xi\left(z\right)^{2}=\Omega_{M}\left(1+z\right)^{3}+\Omega_{K}
\left(1+z\right)^{2}+\Omega_{\Lambda}}$. In this work, we employ the
cosmological "concordance model" of a flat, dark energy and dark matter
dominated Universe with the parameters
$\mathrm{(\Omega_{DM},\Omega_{M},\Omega_{K},\Omega_{\Lambda})}$=$\mathrm{
(0.23,0.27,0,0.73)}$. For the dimensionless Hubble-Parameter h we use the
value 0.71 \cite{WMAP}. The annihilation induced intensity scales
quadratically with the dark matter density and thus strongly depends on the
amount of structure present in the dark matter. This dependence is included
via the function $\mathrm{\Gamma\equiv1/(\overline\rho^{2} V)\;
\int_{V}\rho^{2}dV}$ ( $\mathrm{\overline\rho}$: mean density over volume
V), which we use as z-dependently evaluated for cosmological volumes in
\cite{TaylorandSilk}. Generally speaking, $\mathrm{\Gamma\left(z\right)}$
therefore is the "enhancement factor" between a structured universe and a
completely homogeneous dark matter distribution. This enhancement due to
structure formation is sensitive to the predominant density profile of the
dark matter halos, and therefore subject to some uncertainty.
Most high resolution N-body simulations yield a
universal dark matter halo profile
$\mathrm{\rho(r)=\rho_{S}/[(r/r_{S})^{\gamma}[1+(r/r_{S})^{\gamma}]^{
(\beta-\gamma)/\alpha}}]$, where $\mathrm{\rho_{S}}$ and $\mathrm{r_{S}}$
denote scale density and radius. Mounting evidence of the existence of this
type of dark matter density profile and the validity of the paradigm of
hierarchical structure formation comes from X-ray observations of Abell
clusters \cite{Abell} and from observations of the
Lyman-$\mathrm{\alpha}$-forest at high redshifts \cite{Lyman}. For our
calculations, we will employ the Navarro, Frenk and White (NFW) profile
($\mathrm{\alpha=1,\;\beta=3\;and\;\gamma=1}$) \cite{nfw,nfw2} and a lower
mass cutoff for the halos/subhalos of $\mathrm{10^{5}}$ solar masses as the
baseline case. For the mass-dependent concentration parameter
$\mathrm{c(M_{halo},\,z_{formation})\equiv r_{virial}/r_{S}}$ the results
presented in \cite{ens} are used. This scenario yields a present-day
enhancement of the flux of $\mathrm{2\times 10^{6}}$, compared to a
completely structureless universe \cite{TaylorandSilk}. If a substantial
fraction of the dark matter halos has steeper inner slopes, like the Moore
et al. profile \cite{Moore}, the overall intensity enhancement might well
be a constant factor of 2--25 larger than assumed here (depending on the
inner cutoff-radius in case of a singular inner slope $\mathrm{\propto
r^{-1.5}}$ or steeper) \cite{TaylorandSilk}. Even steeper inner slopes can
arise from adiabatic compression by baryons \cite{Adiabatic}. To account
for this uncertainty, in this paper we will work with the NFW-case of
$\mathrm{\Gamma\left(0\right)=2\times 10^{6}}$, while keeping in mind that
the intensity could be additionally boosted by a factor $\mathrm{\Psi =
\mathcal{O}(1...10)}$. Substantial clumping of the dark matter on mass
scales below $\mathrm{10^{5}}$ solar masses \cite{Diemand} might result 
in further enhancement of the intensities. 
We compare the EGB intensity due to WIMP annihilations with EGRET data
\cite{Strong}, depending on the neutralino parameters $\mathrm{\langle
\sigma v \rangle_{\chi}}$ and $\mathrm{m_{\chi}}$. The EGRET data points in
the energy range 50~MeV -- 300~MeV are very well described by a power law,
presumably due to faint, unresolved active galactic nuclei. The best-fit
spectrum is $\mathrm{7.4\times 10^{-7}\times
(E/GeV)^{-2.33}\,ph\,cm^{-2}\,s^{-1}\,sr^{-1}\,GeV^{-1}}$. A steeper
spectrum than that of the resolved EGRET sources is in fact not unexpected
due to the flux-spectral-index relation \cite{Fluxrel}. Adding the
annihilation spectrum to this steep power law, best fit values for the
cross-section times $\mathrm{\Psi}$ and neutralino mass are
$\mathrm{\langle \sigma v \rangle_{\chi}\times \Psi=(2.6\pm{0.6})\times
10^{-24}\;cm^{3}s^{-1}}$ and $\mathrm{m_{\chi}=515_{-75}^{+110}\;GeV}$
(Fig.~1a). The inferred neutralino mass is independent from the details of
cosmic structure evolution. To verify that correspondingly high values for
$\mathrm{\langle \sigma v \rangle_{\chi}}$ can be obtained within the MSSM
framework while producing cosmologically interesting amounts of
neutralinos, we use the DarkSusy \cite{DarkSusy} numerical routines to scan
the MSSM parameter space. In Fig.~1b, we plot valid models that have been
found in the region of the parameter space described in Table I. In this
$\mathrm{"m_{A}-resonance \; region"}$, annihilation resonantly proceeds
via $\mathrm{\chi\chi\rightarrow A\rightarrow f \overline{f}}$, allowing
for a high annihilation cross-section while still producing
the correct relic density \cite{Baer}. There is considerable spread 
among models. In a number of cases, the observed EGB-signature can
be produced even if 
$\mathrm{\Psi}$ is close or equal to unity.
\begin{table}[tb]
\label{tab:table1}
\caption{Limits of the region of MSSM parameter space that have been
scanned with DarkSusy for cosmologically interesting neutralino models not
excluded by current accelerator limits (higgsino mass parameter
$\mathrm{\mu}$; gaugino mass parameter $\mathrm{m_{2}}$; mass of the cp-odd
higgs $\mathrm{m_{A}}$; ratio of the higgs vacuum expectation values tan
$\mathrm{\beta}$; scalar mass parameter $\mathrm{m_{S}}$ and trilinear
soft-breaking parameters for the third generation squarks $\mathrm{A_{t}}$
and $\mathrm{A_{b}}$)}
\begin{ruledtabular}
\begin{tabular}{cccccccc}$\mathrm{|\mu|}$ &$\mathrm{|m_{2}|}$
&$\mathrm{m_{A}}$ &$\mathrm{tan\, \beta}$ &$\mathrm{m_{S}}$
&$\mathrm{A_{t}}$ &$\mathrm{A_{b}}$\\$\mathrm{500\,GeV}$
&$\mathrm{2500\,GeV}$ &$\mathrm{1000\,GeV}$ &$\mathrm{5}$
&$\mathrm{1000\,GeV}$ &$\mathrm{0.1}$
&$\mathrm{-2.5}$\\$\mathrm{1000\,GeV}$ &$\mathrm{1000\,GeV}$
&$\mathrm{1500\,GeV}$ &$\mathrm{50}$ &$\mathrm{3000\,GeV}$ &$\mathrm{1}$
&$\mathrm{-1}$\\
\end{tabular}
\end{ruledtabular}
\end{table}
\begin{figure}[t]
\includegraphics[width=0.45\textwidth]{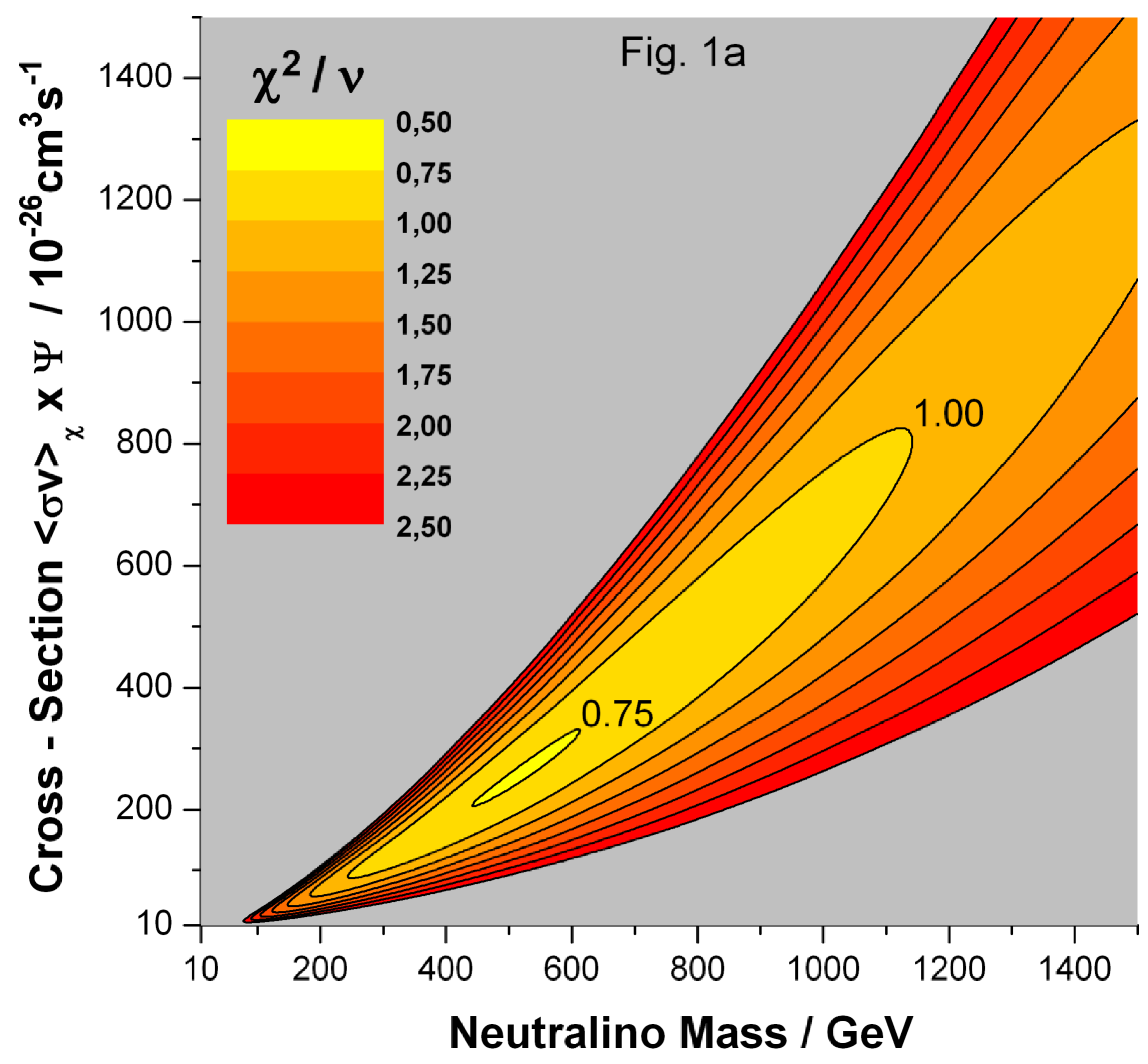}
	\label{fig:egrb1a}
\end{figure}
\begin{figure}[t]
\includegraphics[width=0.45\textwidth]{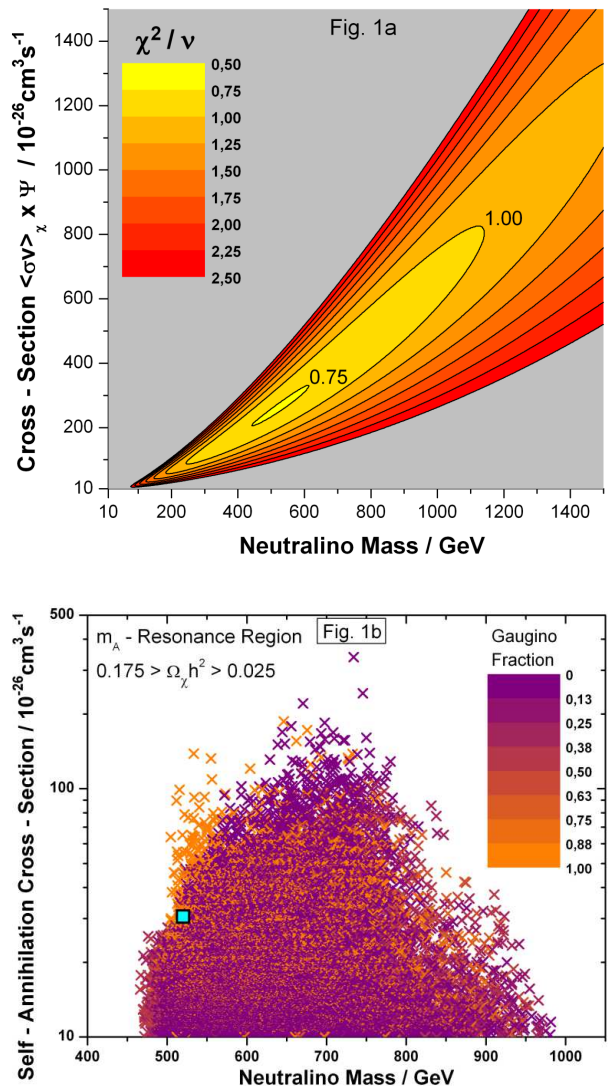}
\caption{(a): Results of the $\mathrm{\chi^{2}}$-test of the neutralino
annihilation hypothesis against the measured EGB. The annihilation
cross-section times $\mathrm{\Psi}$ is normalized to the NFW-profile case
($\mathrm{\Gamma\left(0\right)=2\times 10^{6}}$) \\
(b): Scatter plot of MSSM neutralinos created by scanning the parameter
space described in Table I; the rectangle denotes the 520~GeV neutralino
further explored in Fig.~2}	
\label{fig:egrb1b}
\end{figure}
The models we plot are required to thermally produce $\mathrm{0.175 >
\Omega_{\chi}h^{2} > 0.025}$. For models producing substantially less than
$\mathrm{\Omega_{\chi}h^{2}=0.1}$ an additional, non-thermal source of
neutralinos, e. g. from the decay of heavier relic particles, might be
considered. For a MSSM-neutralino with a mass of 520~GeV, $\mathrm{\langle
\sigma v \rangle_{\chi}=3.1 \times 10^{-25}\,cm^{3}\,s^{-1}}$ and a
moderate $\mathrm{\Psi \;of\; 8}$ the value of $\mathrm{\chi^{2}/\nu}$ is
0.74, which is excellent. The MSSM parameters and resulting EGB spectrum
for this model are shown in Fig.~2. This neutralino is gaugino-like
(gaugino fraction 0.996) and thermally produces the correct relic density
of $\mathrm{\Omega_{\chi}h^{2}\approx 0.1}$. In this scenario the mass of
the lightest Higgs boson $\mathrm{H_{2}}$ is 118~GeV. 
WIMPs with
similar mass and cross-section, but in other respects different 
parameters, might, however, equally be a possibility. 
\begin{figure}[b]	
\begin{center}	
\includegraphics[width=0.45\textwidth]{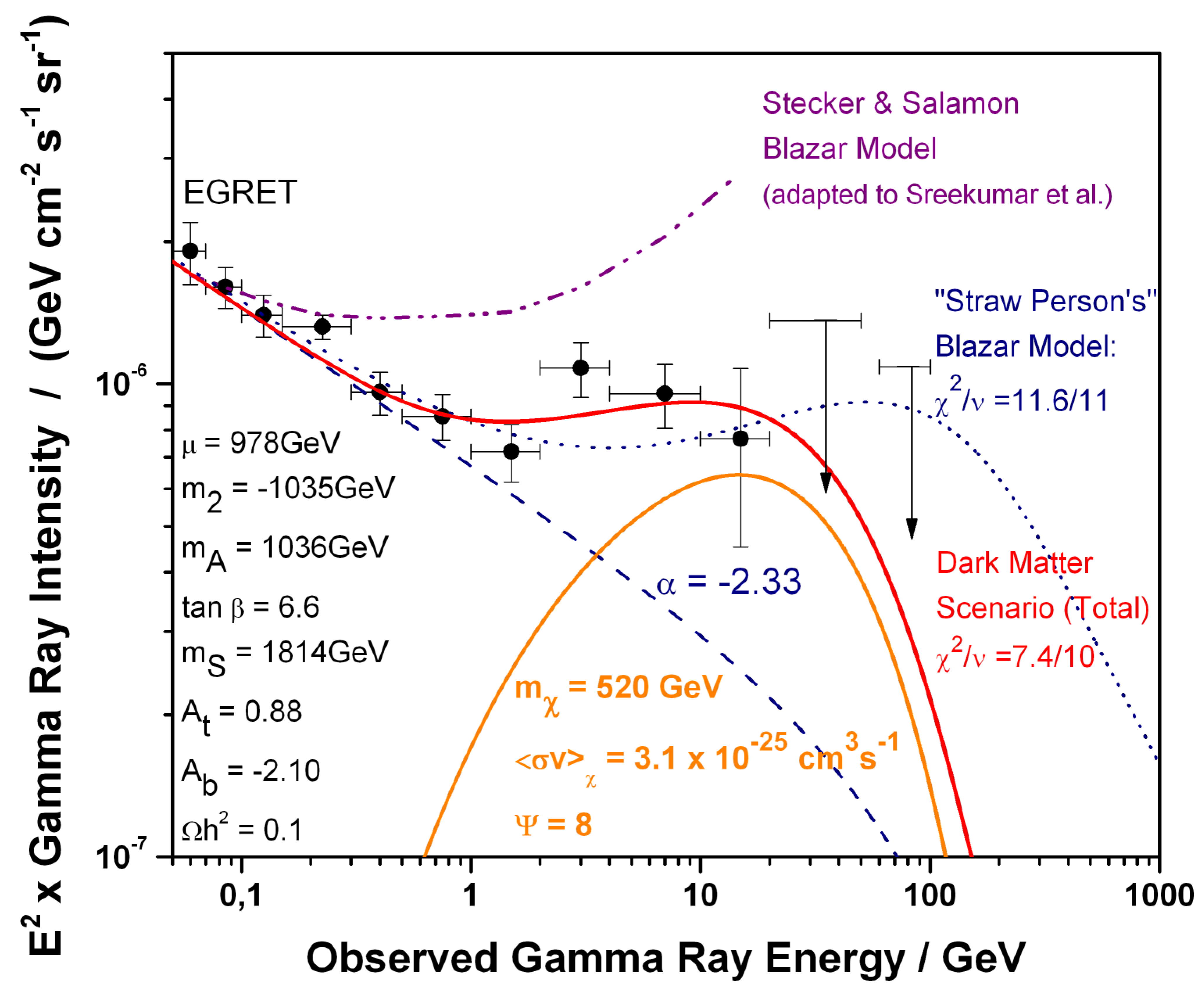}	
\end{center}
\label{fig:egrb2}	
\caption{Extragalactic gamma-ray background: spectrum as determined from
EGRET data by Strong et al.~(data points); the upper limit in the
(60--100)~GeV range is from Sreekumar et al.; steep power law
component~(dashed), Stecker \& Salamon blazar model~(dot-dashed),  straw
person's blazar model~(dotted line), neutralino annihilation
spectrum~(orange solid line), and combined steep power law plus 
annihilation spectrum~(red solid line)}
\end{figure}
\section{Comparison with Astrophysical Background Models} 
In order to explain the weak concave behavior of the EGB intensity above
1~GeV, as it had emerged from the first analysis of the EGRET data
\cite{Sreekumar}, a two component nature of the variable blazar spectra was
assumed: a steep power law as a stationary emission component, and a
flatter power law as a flaring component \cite{SteckerandSalamon}. The
predicted EGB intensity fit the Sreekumar et al. analysis, but does not
agree well with the strong bump feature evident in the new determination of
the EGB (Fig.~2). Constructing a straw persons's model with the same
redshift evolution, we modify the assumptions of Stecker \& Salamon (SS96)
by adopting steeper spectra for the quiescent (faint) component, and adding
a flatter (flaring) spectral component with the hardest spectral index
determined from EGRET data for a single source, to see how well the new
result for the EGB can be matched. Evaluating
$\mathrm{\Phi_\gamma^{AGN}\propto \int dV_{c}\, n(z)
(1+z)^{2}[(E(1+z)/E_{\rm b})^{-2.33}+}$\\$\mathrm{(E(1+z)/E_{\rm
b})^{-1.5}]} \mathrm{\times \kappa(E,z)}$ with the source density in the
co-moving frame $\mathrm{n(z)\propto (1+z)^{3.4}}$ in the redshift range
$\mathrm{0.03\leq z \leq 1.5}$, we obtain a coarse, rescaled version of the
SS96-model, in which the amplitude and break energy $\mathrm{E_{\rm b}}$
were chosen to minimize $\mathrm{\chi^{2}/\nu}$. Details of the luminosity
evolution are unimportant for this test, and the original SS96-curve can be
reproduced accurately by choosing appropriate values for the parameters.
The result of fitting this straw person's model to the new EGB data is a
value of 1.05 (Fig.~2). It should be noted that, while the astrophysical
model can in principle produce an acceptable fit to the data, this requires
a sharp spectral break at an energy $\mathrm{E_{b}}$ of
$\mathrm{\sim5}$~GeV. Blazars, however, have continuously varying spectral
properties (spectral index, peak energies etc.). A sharply bimodal
distribution of the gamma-ray spectral index - as required here - seems
unnatural, and the physical origin of the universal crossing energy thus
mysterious. Moreover, the fraction of sources with hard spectra at an
energy of 1~GeV at any time would have to be about 20\% - considerably more
than the fraction of the hard-spectrum sources in the EGRET catalog
\cite{3EGRET}. The blazar model also would imply $\sim1000$ sources with a
$\mathrm{>300}$~GeV flux of the order of a typical Whipple-source, whereas
the steeper power law alone corresponds to $\sim40$ sources. The first
number seems worryingly large in view of the $\mathrm{\sim10}$ confirmed
sources, in spite of excessive observation campaigns on candidate sources
from radio and X-ray catalogues \cite{HoranWeekes}. Source-intrinsic 
cutoffs well below 100~GeV could, however, remedy this problem. 
The new-generation
Cherenkov telescopes H.E.S.S., MAGIC, VERITAS, and the GLAST observatory
will tell the story.\\ $\mathrm{\emph{Discussion:}\;}$We have arrived at
the conclusion that the neutralino dark matter scenario is in agreement
with the observed EGB spectrum. The best-fit value for the neutralino 
mass
is $\mathrm{m_{\chi}= 515_{-75}^{+110}}$~GeV (notable systematic errors 
of $\mathrm{\sim30\%}$ can be inferred from the systematic uncertainties 
of the EGRET EGB determination \cite{Strong}; they will be
substantially reduced by GLAST). 
The strength of the
observed signature can be explained by a combination of NFW-type dark
matter halo profiles and an annihilation cross-section of the order
that can be obtained within the MSSM framework.
The rather high scale for
the mass ladder of the superpartners might render direct detection of
all but the lightest of these particles by
the LHC difficult \cite{Ellis}. For the present generation of
elastic-scatter experiments, the WIMP-nucleon cross-sections for the
majority of these models are out of reach
($\mathrm{\sigma_{\chi-p}<10^{-7}pb}$), but could be accessible by 
next-generation detectors. For the EGB spectrum to be fitted 
acceptably with
blazar models of the Stecker \& Salamon type, these seem to run into
several worrying difficulties: A universal spectral crossing energy of a
few GeV is required by conspiracy, the models predict a higher fraction of
hard-spectrum sources at GeV energies than observed by EGRET, and the
models possibly imply a higher number of low-redshift blazars 
above 300~GeV than discovered with imaging air Cherenkov telescopes.
The combination of astrophysical sources and WIMP
annihilation presented above is an interesting alternative scenario 
that should be
probed by the next generation of gamma
experiments. 
For the $\mathrm{\sim 520~GeV}$ neutralino
previously discussed, 
the galactic neutralino population would give
rise to a faint galactic gamma-ray halo with intensity 
$\sim10^{-7}$~ph~cm$^{-2}$~s$^{-1}$~sr$^{-1}$ above 1~GeV. 
Present data do not
allow to confirm or rule out such a halo component, but for a galactic
GeV-component commonly attributed to Inverse Compton radiation also 
neutralino annihilation has been proposed as a source \cite{Dixon,DeBoer}. 
A robust calculation using DarkSusy shows that the corresponding 
galactic antiproton flux is not incompatible with the BESS 
measurements \cite{BESS}. If the
astrophysical background can be characterised, gamma rays due to neutralino
annihilation from the Galactic Center or external galaxies like M87
\cite{Baltz} could possibly be detected with low-threshold gamma ray
telescopes. For a NFW-profile and the neutralino presented in Fig. 2, the
Galactic Center would exhibit a gamma ray luminosity of $\mathrm{5\times
10^{35}ergs/s}$ above 1~GeV
$\mathrm{(F_{E>50~GeV}=8\times10^{-11}\,ph\;cm^{-2}\,s^{-1}})$ from within
$\mathrm{10^{-3}sr}$. Note that in this scenario the high EGB intensity 
corresponds to a moderate flux from the Galactic Center 
due to the intricate effects of 
clumping of the dark matter (cf. \cite{SAndo}).
For the MAGIC telescope at an energy threshold of
50~GeV and the halo profile from \cite{Tsai}, the annihilation component in
the gamma ray spectrum of M87 would be detectable at $\mathrm{5\sigma}$
in 250 hours of observation time. --- \small{\textit{We thank A. Strong,
T. Kneiske, M.
Merck and R. R$\ddot{u}$ckl for valuable discussions. Support by BMBF
(O5CM0MG1) and Helmholtz Gemeinschaft (VIHKOS) is gratefully
acknowledged.}}

\begin{thebibliography}{34}
\small{
\expandafter\ifx\csname natexlab\endcsname\relax\def\natexlab#1{#1}\fi
\expandafter\ifx\csname bibnamefont\endcsname\relax
  \def\bibnamefont#1{#1}\fi
\expandafter\ifx\csname bibfnamefont\endcsname\relax
  \def\bibfnamefont#1{#1}\fi
\expandafter\ifx\csname citenamefont\endcsname\relax
  \def\citenamefont#1{#1}\fi
\expandafter\ifx\csname url\endcsname\relax
  \def\url#1{\texttt{#1}}\fi
\expandafter\ifx\csname urlprefix\endcsname\relax\def\urlprefix{URL }\fi
\providecommand{\bibinfo}[2]{#2}
\providecommand{\eprint}[2][]{\url{#2}}

\bibitem[{\citenamefont{Morrison}(1958)}]{Morrison}
\bibinfo{author}{\bibfnamefont{P.}~\bibnamefont{Morrison}},
  \bibinfo{journal}{Nuovo Cimento} \textbf{\bibinfo{volume}{7}},
  \bibinfo{pages}{858} (\bibinfo{year}{1958}).

\bibitem[{\citenamefont{Nolan et~al.}(1992)}]{CGRO}
\bibinfo{author}{\bibfnamefont{P.~L.} \bibnamefont{Nolan}}
  \bibnamefont{et~al.}, \bibinfo{journal}{IEEE Transactions Nucl. Sci.}
  \textbf{\bibinfo{volume}{39}}, \bibinfo{pages}{993}
(\bibinfo{year}{1992}).

\bibitem[{\citenamefont{Dixon et~al.}(1998)}]{Dixon}
\bibinfo{author}{\bibfnamefont{D.~D.} \bibnamefont{Dixon}}
  \bibnamefont{et~al.}, \bibinfo{journal}{New Astronomy}
  \textbf{\bibinfo{volume}{3}}, \bibinfo{pages}{539}
(\bibinfo{year}{1998}).

\bibitem[{\citenamefont{Strong et~al.}(2004)\citenamefont{Strong,
Moskalenko,
  and Reimer}}]{Strong}
\bibinfo{author}{\bibfnamefont{A.~W.} \bibnamefont{Strong}},
  \bibinfo{author}{\bibfnamefont{I.~V.} \bibnamefont{Moskalenko}},
  \bibnamefont{and}
\bibinfo{author}{\bibfnamefont{O.}~\bibnamefont{Reimer}},
\bibinfo{journal}{Astrophys. J.}
  \textbf{\bibinfo{volume}{613}}, \bibinfo{pages}{956}
(\bibinfo{year}{2004}).

\bibitem[{\citenamefont{Sreekumar et~al.}(1998)}]{Sreekumar}
\bibinfo{author}{\bibfnamefont{P.}~\bibnamefont{Sreekumar}}
  \bibnamefont{et~al.}, \bibinfo{journal}{Astrophys. J.}
  \textbf{\bibinfo{volume}{494}}, \bibinfo{pages}{523}
(\bibinfo{year}{1998}).

\bibitem[{\citenamefont{Chiang and Mukherjee}(1998)}]{ChiangMukher}
\bibinfo{author}{\bibfnamefont{J.}~\bibnamefont{Chiang}} \bibnamefont{and}
  \bibinfo{author}{\bibfnamefont{R.}~\bibnamefont{Mukherjee}},
  \bibinfo{journal}{Astrophys. J.} \textbf{\bibinfo{volume}{496}},
  \bibinfo{pages}{752} (\bibinfo{year}{1998}).

\bibitem[{\citenamefont{Stecker and Salamon}(1996)}]{SteckerandSalamon}
\bibinfo{author}{\bibfnamefont{F.~W.} \bibnamefont{Stecker}}
\bibnamefont{and}
  \bibinfo{author}{\bibfnamefont{M.~H.} \bibnamefont{Salamon}},
  \bibinfo{journal}{Astrophys. J.} \textbf{\bibinfo{volume}{464}},
  \bibinfo{pages}{600} (\bibinfo{year}{1996}).

\bibitem[{\citenamefont{Gabici and Blasi}(2003)}]{Blasi}
\bibinfo{author}{\bibfnamefont{S.}~\bibnamefont{Gabici}} \bibnamefont{and}
\bibinfo{author}{\bibfnamefont{P.}~\bibnamefont{Blasi}},
  \bibinfo{journal}{Astropart. Phys.} \textbf{\bibinfo{volume}{19}},
  \bibinfo{pages}{679} (\bibinfo{year}{2003}).

\bibitem[{\citenamefont{Totani}(1999)}]{Totani}
\bibinfo{author}{\bibfnamefont{T.}~\bibnamefont{Totani}},
  \bibinfo{journal}{Astropart. Phys.} \textbf{\bibinfo{volume}{11}},
  \bibinfo{pages}{451} (\bibinfo{year}{1999}).

\bibitem[{\citenamefont{Boehm et~al.}(2004)}]{Boehm}
\bibinfo{author}{\bibfnamefont{C.}~\bibnamefont{Boehm}}
\bibnamefont{et~al.},
  \bibinfo{journal}{Phys. Rev. Lett.} \textbf{\bibinfo{volume}{92}},
  \bibinfo{pages}{101301} (\bibinfo{year}{2004}).

\bibitem[{\citenamefont{Pagels and Primack}(1982)}]{Primack}
\bibinfo{author}{\bibfnamefont{H.}~\bibnamefont{Pagels}} \bibnamefont{and}
  \bibinfo{author}{\bibfnamefont{J.~R.} \bibnamefont{Primack}},
  \bibinfo{journal}{Phys. Rev. Lett.} \textbf{\bibinfo{volume}{48}},
  \bibinfo{pages}{223} (\bibinfo{year}{1982}).

\bibitem[{\citenamefont{G.~Jungman and Griest}(1996)}]{SUSY}
\bibinfo{author}{\bibfnamefont{M.~K.} \bibnamefont{G.~Jungman}}
  \bibnamefont{and}
\bibinfo{author}{\bibfnamefont{K.}~\bibnamefont{Griest}},
  \bibinfo{journal}{Phys. Rept.} \textbf{\bibinfo{volume}{267}},
  \bibinfo{pages}{195} (\bibinfo{year}{1996}).

\bibitem[{\citenamefont{Bergstrom et~al.}(2001)\citenamefont{Bergstrom,
  Edsjo, and Ullio}}]{BEU}
\bibinfo{author}{\bibfnamefont{L.}~\bibnamefont{Bergstrom}},
  \bibinfo{author}{\bibfnamefont{J.}~\bibnamefont{Edsjo}},
\bibnamefont{and}
  \bibinfo{author}{\bibfnamefont{P.}~\bibnamefont{Ullio}},
  \bibinfo{journal}{Phys. Rev. Lett.} \textbf{\bibinfo{volume}{87}},
  \bibinfo{pages}{251301} (\bibinfo{year}{2001}).

\bibitem[{\citenamefont{Ullio et~al.}(2002)\citenamefont{Ullio
et~al.}}]{cl}
\bibinfo{author}{\bibnamefont{P.}~\bibnamefont{Ullio}},
  \bibinfo{author}{\bibnamefont{L.}~\bibnamefont{Bergstrom}},
  \bibinfo{author}{\bibnamefont{J.}~\bibnamefont{Edsjo}} 
  \bibnamefont{and}
  \bibinfo{author}{\bibnamefont{C.}~\bibnamefont{Lacey}},
  \bibinfo{journal}{Phys. Rev. D} \textbf{\bibinfo{volume}{66}},
  \bibinfo{pages}{123502} (\bibinfo{year}{2002}).

\bibitem[{\citenamefont{Gondolo et~al.}(2003{\natexlab{a}})}]{Gondolo}
\bibinfo{author}{\bibfnamefont{P.}~\bibnamefont{Gondolo}}
\bibnamefont{et~al.},
  \bibinfo{journal}{Proc. of the 4th IDM, York, UK} p. 
  \bibinfo{pages}{256}
  (\bibinfo{year}{2003}{\natexlab{a}}).

\bibitem[{\citenamefont{Elsaesser and Mannheim}(2004)}]{Elsaesser1}
\bibinfo{author}{\bibfnamefont{D.}~\bibnamefont{Elsaesser}}
\bibnamefont{and}
  \bibinfo{author}{\bibfnamefont{K.}~\bibnamefont{Mannheim}},
  \bibinfo{journal}{Astropart. Phys.} \textbf{\bibinfo{volume}{22}},
  \bibinfo{pages}{65} (\bibinfo{year}{2004}).

\bibitem[{\citenamefont{Kneiske et~al.}(2004)}]{Kneiske}
\bibinfo{author}{\bibfnamefont{T.}~\bibnamefont{Kneiske}}
\bibnamefont{et~al.},
  \bibinfo{journal}{Astron. Astrophys.} 
  \textbf{\bibinfo{volume}{413}},
  \bibinfo{pages}{807} (\bibinfo{year}{2004}).

\bibitem[{\citenamefont{Zdziarski and Svensson}(1989)}]{Zdziarski}
\bibinfo{author}{\bibfnamefont{A.~A.} \bibnamefont{Zdziarski}}
  \bibnamefont{and}
\bibinfo{author}{\bibfnamefont{R.}~\bibnamefont{Svensson}},
  \bibinfo{journal}{Astrophys. J.} \textbf{\bibinfo{volume}{344}},
  \bibinfo{pages}{551} (\bibinfo{year}{1989}).

\bibitem[{\citenamefont{Bennett et~al.}(2003)}]{WMAP}
\bibinfo{author}{\bibfnamefont{C.~L.} \bibnamefont{Bennett}}
  \bibnamefont{et~al.}, \bibinfo{journal}{ApJS}
\textbf{\bibinfo{volume}{148}},
  \bibinfo{pages}{1} (\bibinfo{year}{2003}).

\bibitem[{\citenamefont{Taylor and Silk}(2003)}]{TaylorandSilk}
\bibinfo{author}{\bibfnamefont{J.}~\bibnamefont{Taylor}} 
\bibnamefont{and}
  \bibinfo{author}{\bibfnamefont{J.}~\bibnamefont{Silk}},
  \bibinfo{journal}{MNRAS} \textbf{\bibinfo{volume}{339}},
\bibinfo{pages}{505}
  (\bibinfo{year}{2003}).

\bibitem[{\citenamefont{Lewis et~al.}(2003)
\citenamefont{Lewis, Buote, and
  Stocke}}]{Abell}
\bibinfo{author}{\bibfnamefont{A.~D.} \bibnamefont{Lewis}},
  \bibinfo{author}{\bibfnamefont{D.~A.} \bibnamefont{Buote}},
\bibnamefont{and}
  \bibinfo{author}{\bibfnamefont{J.~T.} \bibnamefont{Stocke}},
  \bibinfo{journal}{Astrophys. J.} \textbf{\bibinfo{volume}{586}},
  \bibinfo{pages}{135} (\bibinfo{year}{2003}).

\bibitem[{\citenamefont{Croft et~al.}(1999)}]{Lyman}
\bibinfo{author}{\bibfnamefont{R.~A.~C.} \bibnamefont{Croft}}
  \bibnamefont{et~al.}, \bibinfo{journal}{Astrophys. J.}
  \textbf{\bibinfo{volume}{520}}, \bibinfo{pages}{1}
(\bibinfo{year}{1999}).

  \bibitem[{\citenamefont{Navarro et~al.}(1996)\citenamefont{Navarro,
Frenk, and
  White}}]{nfw}
\bibinfo{author}{\bibfnamefont{J.~F.} \bibnamefont{Navarro}},
  \bibinfo{author}{\bibfnamefont{C.~S.} \bibnamefont{Frenk}},
\bibnamefont{and}
  \bibinfo{author}{\bibfnamefont{S.~D.~M.} \bibnamefont{White}},
  \bibinfo{journal}{Astrophys. J.} \textbf{\bibinfo{volume}{462}},
  \bibinfo{pages}{563} (\bibinfo{year}{1996}).

\bibitem[{\citenamefont{Navarro et~al.}(1997)
\citenamefont{Navarro, Frenk,
and
  White 2}}]{nfw2}
\bibinfo{author}{\bibfnamefont{J.~F.} \bibnamefont{Navarro}},
  \bibinfo{author}{\bibfnamefont{C.~S.} \bibnamefont{Frenk}},
\bibnamefont{and}
  \bibinfo{author}{\bibfnamefont{S.~D.~M.} \bibnamefont{White}},
  \bibinfo{journal}{Astrophys. J.} \textbf{\bibinfo{volume}{490}},
  \bibinfo{pages}{493} (\bibinfo{year}{1997}).

  \bibitem[{\citenamefont{Eke, Navarro, Steinmetz}(2001)
  \citenamefont{Eke,
Navarro, Steinmetz}}]{ens}
\bibinfo{author}{\bibfnamefont{V.~R.} \bibnamefont{Eke}},
  \bibinfo{author}{\bibfnamefont{J.~F.} \bibnamefont{Navarro}},
\bibnamefont{and}
  \bibinfo{author}{\bibfnamefont{M.} \bibnamefont{Steinmetz}},
  \bibinfo{journal}{Astrophys. J.} \textbf{\bibinfo{volume}{554}},
  \bibinfo{pages}{114} (\bibinfo{year}{2001}).

\bibitem[{\citenamefont{Moore et~al.}(1998)}]{Moore}
\bibinfo{author}{\bibfnamefont{B.}~\bibnamefont{Moore}}
\bibnamefont{et~al.},
  \bibinfo{journal}{Astrophys. J.} \textbf{\bibinfo{volume}{499}},
  \bibinfo{pages}{L5} (\bibinfo{year}{1998}).
  
\bibitem[{\citenamefont{Prada et~al.}(2004)}]{Adiabatic}
\bibinfo{author}{\bibfnamefont{F.}~\bibnamefont{Prada}}
\bibnamefont{et~al.},
  \bibinfo{journal}{Phys. Rev. Lett.} \textbf{\bibinfo{volume}{93}},
  \bibinfo{pages}{241301} (\bibinfo{year}{2004}).

   \bibitem[{\citenamefont{Diemand, Moore, Stadel}(2005)
   \citenamefont{Diemand, Moore, Stadel}}]{Diemand}
\bibinfo{author}{\bibfnamefont{J.} \bibnamefont{Diemand}},
  \bibinfo{author}{\bibfnamefont{B.} \bibnamefont{Moore}},
\bibnamefont{and}
  \bibinfo{author}{\bibfnamefont{J.} \bibnamefont{Stadel}},
  \bibinfo{journal}{Nature} \textbf{\bibinfo{volume}{433}},
  \bibinfo{pages}{389} (\bibinfo{year}{2005}).
   
  \bibitem[{\citenamefont{von~Montigni et~al.}(1997)}]{Fluxrel}
\bibinfo{author}{\bibfnamefont{C.}~\bibnamefont{von~Montigni}}
\bibnamefont{et~al.},
  \bibinfo{journal}{Astrophys. J.} \textbf{\bibinfo{volume}{483}},
  \bibinfo{pages}{161} (\bibinfo{year}{1997}).

\bibitem[{\citenamefont{Gondolo et~al.}(2003){\natexlab{b}})}]{DarkSusy}
\bibinfo{author}{\bibfnamefont{P.}~\bibnamefont{Gondolo}}
\bibnamefont{et~al.},
  \bibinfo{journal}{Proc. of the 4th IDM, York, UK}
  (\bibinfo{year}{2003}{\natexlab{b}}).

\bibitem[{\citenamefont{Baer et~al.}(2004)}]{Baer}
\bibinfo{author}{\bibfnamefont{H.}~\bibnamefont{Baer}}
  \bibnamefont{et~al.}, \bibinfo{journal}{JCAP}
  \textbf{\bibinfo{volume}{0408}}, \bibinfo{pages}{005}
(\bibinfo{year}{2004}).
  
\bibitem[{\citenamefont{Hartman et~al.}(1999)}]{3EGRET}
\bibinfo{author}{\bibfnamefont{R.~C.} \bibnamefont{Hartman}}
  \bibnamefont{et~al.}, \bibinfo{journal}{ApJS}
\textbf{\bibinfo{volume}{123}},
  \bibinfo{pages}{79} (\bibinfo{year}{1999}).

\bibitem[{\citenamefont{Horan and Weekes}(2004)}]{HoranWeekes}
\bibinfo{author}{\bibfnamefont{D.}~\bibnamefont{Horan}} \bibnamefont{and}
  \bibinfo{author}{\bibfnamefont{T.~C.} \bibnamefont{Weekes}},
  \bibinfo{journal}{New Astron. Rev.} \textbf{\bibinfo{volume}{48}},
  \bibinfo{pages}{527} (\bibinfo{year}{2004}).

\bibitem[{\citenamefont{Battaglia et~al.}(2004)}]{Ellis}
\bibinfo{author}{\bibfnamefont{M.}~\bibnamefont{Battaglia}}
  \bibnamefont{et~al.}, \bibinfo{journal}{Euro. Phys. J. C}
  \textbf{\bibinfo{volume}{33}}, \bibinfo{pages}{273}
(\bibinfo{year}{2004}).
  
  \bibitem[{\citenamefont{Aloisio and Blasi and Olinto}(2004)}]{OBGC}
\bibinfo{author}{\bibfnamefont{R.} \bibnamefont{Aloisio}},
  \bibinfo{author}{\bibfnamefont{P.}
  \bibnamefont{Blasi}} \bibnamefont{and}
\bibinfo{author}{\bibfnamefont{A.~V.}
  \bibnamefont{Olinto}}, \bibinfo{journal}{JCAP}
  \textbf{\bibinfo{volume}{0405}}, \bibinfo{pages}{007}
(\bibinfo{year}{2004}).

\bibitem[{\citenamefont{de~Boer et~al.}(2003)}]{DeBoer}
\bibinfo{author}{\bibfnamefont{W.}~\bibnamefont{de~Boer}}
\bibnamefont{et~al.}
  (\bibinfo{year}{2003}), \eprint{hep-ph/0309029}.

\bibitem[{\citenamefont{Matsunaga et~al.}(1998)}]{BESS}
\bibinfo{author}{\bibfnamefont{H.}~\bibnamefont{Matsunaga}}
  \bibnamefont{et~al.}, \bibinfo{journal}{Phys. Rev. Lett.}
  \textbf{\bibinfo{volume}{81}}, \bibinfo{pages}{4052}
(\bibinfo{year}{1998}).

\bibitem[{\citenamefont{Baltz et~al.}(1999)}]{Baltz}
\bibinfo{author}{\bibfnamefont{E.~A.} \bibnamefont{Baltz}}
  \bibnamefont{et~al.}, \bibinfo{journal}{Phys. Rev. D}
  \textbf{\bibinfo{volume}{61}}, \bibinfo{pages}{023514}
(\bibinfo{year}{1999}).

\bibitem[{\citenamefont{S.~Ando}(2005)}]{SAndo}
\bibinfo{author}{\bibfnamefont{S.}~\bibnamefont{Ando}}
  (\bibinfo{year}{2005}), \eprint{astro-ph/0503006}.

\bibitem[{\citenamefont{Tsai}(1993)}]{Tsai}
\bibinfo{author}{\bibfnamefont{J.~C.} \bibnamefont{Tsai}},
  \bibinfo{journal}{Astrophys. J.} \textbf{\bibinfo{volume}{413}},
  \bibinfo{pages}{L59} (\bibinfo{year}{1993}).}

\end{thebibliography}

\end{document}